\documentclass[twocolumn,showpacs,preprintnumbers,amsmath,amssymb,superscriptaddress,floatfix,nofootinbib]{revtex4}
\usepackage{graphicx}
\usepackage{amsmath}
\usepackage{amsfonts}
\usepackage{amssymb}
\usepackage{color}
\usepackage{multirow}
\usepackage{footnote}

\begin{document}
\title{Evasion of HSR in the charmless decays of excited $P$-wave charmonia}

\author{Yan Wang}
\affiliation{College of Physics and Engineering, Qufu Normal
University, Qufu 273165, China}
\author{Qi Wu}
\affiliation{School of Physics, Southeast University, Nanjing
210094, China}

\author{Gang Li} \email{gli@qfnu.edu.cn}
\affiliation{College of Physics and Engineering, Qufu Normal
University, Qufu 273165, China}

\author{Ju-Jun Xie} \email{xiejujun@impcas.ac.cn} \affiliation{Institute of Modern
Physics, Chinese Academy of Sciences, Lanzhou 730000, China}
\affiliation{School of Nuclear Sciences and Technology, University
of Chinese Academy of Sciences, Beijing 101408, China}
\affiliation{School of Physics and Microelectronics, Zhengzhou
University, Zhengzhou, Henan 450001, China}

\author{Chun-Sheng An} \email{ancs@swu.edu.cn}
\affiliation{School of Physical Science and Technology, Southwest
University, Chongqing 400715, China}

\begin{abstract}
\begin{center}
\textbf{Abstract}
\end{center}

We investigate the charmless decays of excited $P$-wave charmonia
$\chi_{c1}^\prime \to VV$ and $\chi_{c2}^\prime \to VP$ via
intermediate charmed meson loops, where $V$ and $P$ denote the light
vector and pseudoscalar mesons, respectively. Within the model
parameters, the charmed meson loop contributions are evaluated by
using the effective Lagrangian approach. For $\chi_{c1}^\prime \to
VV$, the partial widths of the $\rho\rho$, $\omega\omega$, and
$K^*{\bar K}^*$ channels can reach to the order of MeV, while the
partial width of the $\phi\phi$ channel is very small and in the
order of keV. For $\chi_{c2}^\prime \to V P$, the partial widths of
$\chi_{c2}^{\prime} \to K^\ast \bar{K}+c.c$ turns out to be
sizeable, while the partial widths of $\chi_{c2}^{\prime} \to
\rho^+\pi^- +c.c$ is found to be much smaller than the $K^\ast
\bar{K}+c.c$ channel. Our calculations may be examined by the future
BESIII experiments.

\end{abstract}
\date{\today}

\pacs{13.25.GV, 13.75.Lb, 14.40.Pq}
\maketitle
\section{Introduction} \label{sec:introduction}

The energy region of charmonium contains rich information about both
perturbative and nonperturbative QCD dynamics. By studying the
exclusive decays of charmonium, we expect to obtain some insights
into the QCD properties in this regime. The charmless decay modes of
charmonium states are suppressed according to the Okubo-Zweig-Iizuka
(OZI) rule. But, these charmless decays are crucial to understand
the dynamical properties of QCD. For example, in the perturbative
QCD (pQCD) approach, if one considers only the valence Fock state
$c\bar{c}$, the branching ratios of the charmless decay $J_{c\bar
c}(\lambda) \to h_1(\lambda_1) h_2(\lambda_2)$ can be written
as~\cite{Chernyak:1981zz},
\begin{eqnarray}
{\mbox {BR}}[{J_{c\bar c}(\lambda)} \to h_1(\lambda_1) h_2(\lambda_2)] \sim \left( \frac {\Lambda_{QCD}^2} {m_c^2}\right)^{|\lambda_1 + \lambda_2| +2} \, , \label{eq:HSR}
\end{eqnarray}
where $J_{c \bar c}$, $h_1$ and $h_2$ are the initial charmonium
meson and final two light mesons, respectively. $\lambda$,
$\lambda_1$, and $\lambda_2$ are the helicities of the corresponding
mesons. From Eq.~(\ref{eq:HSR}), one can see that the leading
contribution corresponds to the $\lambda_1+\lambda_2 = 0$ condition,
while the helicity configurations that do not satisfy this relation
will be suppressed. This is the so-called helicity selection rule
(HSR).

This HSR can alternatively be described with the ``naturalness"
quantum number of relevant particle $\sigma \equiv P(-1)^J$, where
$P$ and $J$ are the parity and spin of the particle, respectively.
The HSR then requires that $\sigma^{\rm{initial}} =
\sigma_1\sigma_2$, which means that the naturalness of initial state
equals to the product of the final
states~\cite{Chernyak:1981zz,Chernyak:1983ej,Feldmann:2000hs}. If
$\sigma^{\rm{initial}}\neq \sigma_1\sigma_2$, one have to add a
Levi-Civita (LC) tensor $\varepsilon_{\alpha\beta\mu\nu}$ in the
amplitude to keep the parity conservation and Lorentz invariance.
The LC tensor are contracted with the polarization vectors and
momenta of the involved mesons, hence there are some terms vanished
in the helicity amplitudes, and these contributions are supposed to
be suppressed by pQCD.

On the other hand, intermediate meson loop (IML) is regarded as an
important nonperturbative transition mechanism in the charmonium
energy
region~\cite{Lipkin:1988tg,Moxhay:1988ri,Lipkin:1986bi,Lipkin:1986av}.
Recently, this mechanism has been successfully applied to study the
production and decays of charmonium and exotic
states~\cite{Liu:2013vfa,Guo:2013zbw,Wang:2013hga,Cleven:2013sq,Chen:2011pv,Li:2012as,Li:2013yla,Voloshin:2013ez,Voloshin:2011qa,Bondar:2011ev,Chen:2011pu,Chen:2012yr,Chen:2013bha,Li:2015uwa,Li:2014gxa,Li:2014uia,Li:2013zcr,Li:2011ssa,Guo:2010ak,Wu:2016ypc,Wu:2016dws,Liu:2016xly,Li:2014pfa,Yuan-Jiang:2010cna,Zhao:2013jza,Li:2013xia,Li:2007xr,Qin:2019ybr,Liu:2019dqc,Wu:2019vbk,Zhang:2018eeo}.
Some exclusive decay modes of charmonia below the open $D\bar D$
threshold will be suppressed by both OZI rule and HSR and there
shows significant discrepancies between the experimental
measurements and the theoretical
expectations~\cite{Patrignani:2016xqp}. In the previous
works~\cite{Liu:2009vv,Liu:2010um,Wang:2012wj,Li:2013jma,Wang:2012mf},
the HSR violating processes of the charmonium states decaying into
the light vector mesons, pseudoscalar mesons, or baryon-anti-baryon
pairs were studied. The results indicate that the IML transitions
are important to reproduce the experimental data on these decays,
although there are still some model-dependent parameters needed to
be determined with more accurate data. To give a consistent
theoretical description and search for the underlying dynamic
mechanism, investigating many other pertinent HSR violating decay
modes turns to be necessary.

In Ref.~\cite{Liu:2009vv}, the $P$-wave ground states HSR suppressed
decays $\chi_{c1} \to VV$ and $\chi_{c2}\to VP$ were studied via
intermediate charmed meson loops. In this paper, we will further
investigate the excited $P$-wave states HSR suppressed decays
$\chi_{c1}^\prime \to VV$ and $\chi_{c2}^\prime \to VP$~\footnote{We use $\chi_{c1}/\chi_{c2}$ and $\chi_{c1}^\prime/\chi_{c2}^\prime$ for
$\chi_{c1}(1P)/\chi_{c2}(1P)$ and
$\chi_{c1}(2P)/\chi_{c2}(2P)$, respectively.}. Since the
$\chi_{c1}^\prime$ and $\chi_{c2}^\prime$ are above the open charmed meson
pairs, it is expected that the IML mechanims should be more
important in the above HSR violating processes.

The paper is organized as follows. We present our model and
ingredients of the effective Lagrangians and decay amplitudes in
Sec.~\ref{sec:The model}. The numerical results are shown in
Sec.~\ref{sec:numerical-results} and the summary is presented in
Sec.~\ref{sec:summary}.

\section{The model} \label{sec:The model}

\begin{figure}[htbp]
\centering
\includegraphics[scale=1.0]{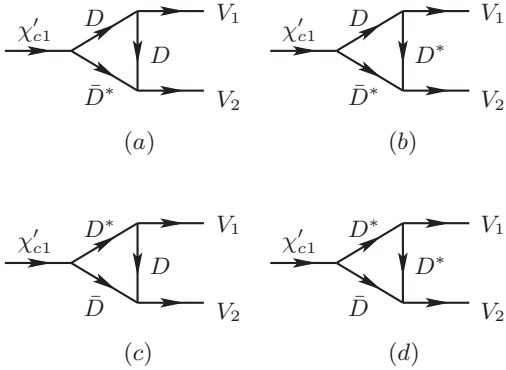}
\caption{The hadron-level diagrams for charmless decay modes
$\chi_{c1}^{\prime} \to VV$ via intermediate charmed meson loops.}
\label{fig:feyn_chic1}
\end{figure}

\begin{figure}[ht]
\centering
\includegraphics[scale=1.0]{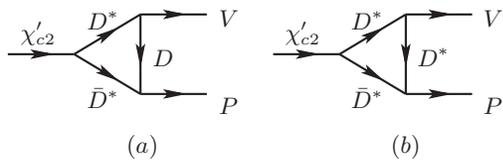}
\caption{The hadron-level diagrams for charmless decay modes
$\chi_{c2}^{\prime} \to VP$ via intermediate charmed meson loops.}
\label{fig:feyn_chic2}
\end{figure}

Following Ref.~\cite{Liu:2009vv}, we consider the contributions of
intermediate meson loops as illustrated in
Figs.~\ref{fig:feyn_chic1} and~\ref{fig:feyn_chic2} for the decays
of $\chi_{c1}^{\prime} \to VV$ and $\chi_{c2}^{\prime} \to VP$,
respectively. In fact, we should also take other possible
intermediate meson loops into account. Since $\chi_{cJ}^{\prime}$
couple to two charmed mesons in $S$-wave and the masses of
$\chi_{cJ}^{\prime}$ are near the mass threshold of the charmed
mesons pairs, we consider the IML illustrated in
Figs.~\ref{fig:feyn_chic1} and ~\ref{fig:feyn_chic2} as the leading
order contributions for the $\chi_{c1}^{\prime} \to VV$ and
$\chi_{c2}^{\prime} \to VP$ decays. Note that $\chi'_{c2}$ couples
to $D\bar{D}$ in $D$-wave, however, this contribution is much
smaller compared with the contributions shown in
Fig.~\ref{fig:feyn_chic2}, and they are safely neglected.

To calculate the decay amplitudes of these diagrams, we need the
effective interaction Lagrangians for each vertex of
Figs.~\ref{fig:feyn_chic1} and~\ref{fig:feyn_chic2}. Based on the
heavy quark limit and chiral
symmetry~\cite{Casalbuoni:1996pg,Colangelo:2003sa,Cheng:2004ru}, the
Lagrangian for the $P$-wave charmonia at leading order is given by
\begin{eqnarray}
\mathcal L_1&=&
i{g_1}Tr\left[{P^\mu_{c \bar c}}{\bar H_{2i}}\gamma^\mu{\bar H_{1i}} \right ]+h.c.,
\label{Lagrangian P}
\end{eqnarray}
where the spin multiplets for these $P$-wave charmonium states are
expressed as:
\begin{eqnarray}
P^\mu_{c \bar c} &=& \frac{1 + \not
v}{2}\left(\chi^{\mu\alpha}_{c2}\gamma_{\alpha}+\frac{1}{\sqrt{2}}\epsilon_{\mu\nu\alpha\beta}v^{\alpha}\gamma^{\beta}\chi^{\nu}_{c1}
\right.
\nonumber \\
&& \left. +\frac{1}{\sqrt{3}}(\gamma^{\mu}-v^{\mu})\chi_{c0}
+h^\mu_c\gamma_5 \right)\frac{1 - \not v}{2} \, ,
\end{eqnarray}
with $v^\mu$ being the four-velocity of the multiplets. Besides, the
charmed and anti-charmed mesons triplets read as:
\begin{eqnarray}
H_{1i} &=& \frac{1 + \not v}{2}\left[{\cal D}_{i\mu}^* \gamma_\mu - {\cal D}_i \gamma_5 \right],\nonumber\\
H_{2i} &=& \left[ {\bar {\cal D}}_{i\mu}^*\gamma_\mu + {\bar {\cal D}}_i\gamma_5 \right]\frac{1 - \not v}{2},
\end{eqnarray}
where ${\cal D}$ and ${\cal D}^*$ denote the pseudoscalar and vector
charmed meson fields, respectively, i.e. ${\cal D}^{(*)} =
(D^{(*)+}, D^{(*)0}, D^{(*)+}_{s})$. $v^\mu$ is the four-velocity of
the intermediate charmed mesons. $\varepsilon_{\mu\nu\alpha\beta}$
is the antisymmetric LC tensor and $\varepsilon_{0123}=+1$.

Consequently, the explicit Lagrangian of $P$-wave charmonium
$\chi_{cJ}$ is expressed as,
\begin{eqnarray}
\mathcal{L}_{P}&=& i g_{\chi_{c0} {\cal D} {\cal D}} {\cal D}^i
{\cal D}_i^\dag + i g_{\chi_{c0} {\cal D}^* {\cal D}^*} {\cal
D}^{*i}_\mu {\cal D}_i^{*\mu\dag} \nonumber \\
&& +g_{\chi_{c1} {\cal D}^{*} {\cal D}} \chi_{c1}^\mu ({\cal
D}^*_{i\mu} {\cal D}^{i\dag}+{\cal D}_i {\cal D}^{i\mu\dag} )
\nonumber \\ && +i g_{\chi_{c2} {\cal D}^{*} {\cal D}^{*}}
\chi_{c2}^{\alpha \beta} ({\cal D}_{\alpha}^{i*} {\cal
D}_{i\beta}^{* \dagger}+ {\cal D}_{\beta}^{i*} {\cal D}_{i\alpha}^{*
\dagger}) \, ,
\end{eqnarray}
where the coupling constants will be discussed in the following.

In addition, the Lagrangians relevant to the light vector and
pseudoscalar mesons can be constructed based on the heavy quark
limit and chiral symmetry,
\begin{eqnarray}
{\cal L} &=& -ig_{{\cal D}^{\ast }{\cal D} {\mathcal P}}\left( {\cal
D}^i \partial^\mu {\mathcal P}_{ij} {\cal D}_\mu^{\ast j\dagger
}-{\cal D}_\mu^{\ast i}\partial^\mu {\mathcal
P}_{ij} {\cal D}^{j \dag}\right) \nonumber \\
&& +\frac{1}{2}g_{{\cal D}^\ast D^\ast {\mathcal P}}\varepsilon
_{\mu \nu \alpha \beta }{\cal D}_i^{\ast \mu }\partial^\nu {\mathcal
P}^{ij}  {\overset{ \leftrightarrow }{\partial }}{\!^{\alpha }}
{\cal D}_j^{\ast \beta\dag } \nonumber
\\
&& -2f_{\mathcal{D}^*\mathcal{D}\mathcal{V}}
\epsilon_{\mu\nu\alpha\beta} (\partial^\mu \mathcal{V}^\nu)^i_j
(\mathcal{D}_i^\dagger{\stackrel{\leftrightarrow}{\partial}}{\!^\alpha}
\mathcal{D}^{*\beta
j}-\mathcal{D}_i^{*\beta\dagger}{\stackrel{\leftrightarrow}{\partial}}{\!^\alpha}
{\cal D}^j) \nonumber
\\
&&  - ig_{\mathcal{D}\mathcal{D}\mathcal{V}} \mathcal{D}_i^\dagger
{\stackrel{\leftrightarrow}{\partial}}{\!_\mu}
\mathcal{D}^j(\mathcal{V}^\mu)^i_j
+ ig_{\mathcal{D}^*\mathcal{D}^*\mathcal{V}}
\mathcal{D}^{*\nu\dagger}_i
{\stackrel{\leftrightarrow}{\partial}}{\!_\mu}
\mathcal{D}^{*j}_\nu(\mathcal{V}^\mu)^i_j \nonumber
\\
&& +4if_{\mathcal{D}^*\mathcal{D}^*\mathcal{V}}
\mathcal{D}^{*\dagger}_{i\mu}(\partial^\mu
\mathcal{V}^\nu-\partial^\nu \mathcal{V}^\mu)^i_j
\mathcal{D}^{*j}_\nu +{\rm H.c.} , \label{eq:light-meson}
 \label{eq:LDDV}
 \end{eqnarray}
where $\mathcal P$ and ${\mathcal V}_\mu$ are $3\times 3$ matrices
for the octet pseudoscalar and nonet vector mesons, respectively,
 \begin{eqnarray}  \mathcal{P} &= &\left(
\begin{array} {ccc}
 \frac {\pi^0} {\sqrt {2}} +\frac {\eta} {\sqrt{6}} & \pi^+ & K^+ \\
\pi^- & -\frac {\pi^0} {\sqrt {2}} +\frac { \eta} {\sqrt{6}} & K^0 \\
K^-& {\bar K}^0 & -\sqrt{\frac{2}{3}}\eta \\
\end{array}\right), \\
\mathcal{V}_\mu &=& \left(\begin{array}{ccc}\frac{\rho^0} {\sqrt {2}}+\frac {\omega} {\sqrt {2}}&\rho^+ & K^{*+} \\
\rho^- & -\frac {\rho^0} {\sqrt {2}} + \frac {\omega} {\sqrt {2}} & K^{*0} \\
K^{*-}& {\bar K}^{*0} & \phi \\
\end{array}\right)_\mu \, .
\end{eqnarray}

Then we can write the explicit transition amplitudes for
$\chi_{c1}^{\prime}(p_1)\to [D^{(*)}(q_1) {\bar D}^{(*)}(q_3)]
D^{(*)}(q_2) \to V_1(p_2)V_2(p_3)$ as follows:

\begin{eqnarray}
\mathcal{M}_a&=& \int\frac{d^4q_2}{(2\pi)^4} [g_{\chi_{c1}^{\prime}D^*D}\epsilon_{1\alpha}][-g_{DDV} (q_1+q_2)_\mu\epsilon_2^{*\mu}] \nonumber
\\
&& \times [2f_{D^*DV} \varepsilon_{\kappa \lambda \rho\sigma} p_3^\kappa \epsilon_3^{*\lambda} (q_2-q_3)^\rho ] \frac {i} {q_1^2-m_1^2}\nonumber\\
&&\times \frac {i}
 {q_2^2-m_2^2} \frac {i(-g^{\alpha\sigma}+{q^\alpha_3q^\sigma_3}/{m_3^2})} {q_3^2-m_3^2} {\cal F}(q^2), \nonumber \\
\mathcal{M}_b&=& \int\frac{d^4q_2}{(2\pi)^4} [g_{\chi_{c1}^{\prime}D^*D}\epsilon_{1\alpha}][2f_{D^*DV} \varepsilon_{\mu\nu\theta\phi} (q_1+q_2)^\theta \nonumber \\
&& \times p_2^\mu \epsilon_2^{*\phi} ] [g_{D^*D^*V} (q_2-q_3)_\rho g_{\lambda\sigma}\epsilon_3^{*\rho} -4f_{D^*D^*V} \nonumber \\ && \times (p_{3\sigma}g_{\lambda\rho}-p_{3\lambda}g_{\sigma\rho})\epsilon_3^{*\rho}]\frac {i} {q_1^2-m_1^2} \nonumber\\
&&\times \frac {i(-g^{\nu\lambda}+{q^\nu_2q^\lambda_2}/{m_2^2})}
 {q_2^2-m_2^2}\frac {i(-g^{\alpha\sigma}+{q^\alpha_3q^\sigma_3}/{m_3^2})} {q_3^2-m_3^2} {\cal F}(q^2), \nonumber \\
\mathcal{M}_c&=& \int\frac{d^4q_2}{(2\pi)^4} [g_{\chi_{c1}^{\prime}D^*D}\epsilon_{1\alpha}][-2f_{D^*DV} \varepsilon_{\mu\nu\theta\phi} (q_1+q_2)^\theta ]\nonumber\\
&&\times p_2^\mu \epsilon_2^{*\nu} [-g_{DDV}(q_2-q_3)_\kappa \epsilon_3^{*\kappa}] \frac {i(-g^{\alpha\phi}+{q^\alpha_1q^\phi_1}/{m_1^2})} {q_1^2-m_1^2} \nonumber \\
&& \times \frac {i}
 {q_2^2-m_2^2} \frac {i} {q_3^2-m_3^2} {\cal F}(q^2), \nonumber \\
\mathcal{M}_d&=& \int\frac{d^4q_2}{(2\pi)^4} [g_{\chi_{c1}^{\prime}D^*D}\epsilon_{1\alpha}] [g_{D^*D^*V} (q_1+q_2)_\phi g_{\mu\theta}\epsilon_2^{*\phi} \nonumber \\
&& -4f_{D^*D^*V}(p_{2\theta}g_{\mu\phi}-p_{2\mu}g_{\theta\phi})\epsilon_2^{*\phi}][2f_{D^*DV} \varepsilon_{\kappa\lambda\rho\sigma} \nonumber\\
&&\times
p_3^\kappa\epsilon_3^{*\lambda} (q_2-q_3)^\rho ] \frac {i(-g^{\alpha\mu}+{q^\alpha_1q^\mu_1}/{m_1^2})} {q_1^2-m_1^2}\nonumber \\
&& \times \frac {i(-g^{\theta\sigma}+{q^\theta_2q^\sigma_2}/{m_2^2})}
 {q_2^2-m_2^2} \frac {i} {q_3^2-m_3^2} {\cal F}(q^2) \, ,
\end{eqnarray}

where $p_1$ ($\varepsilon_1$), $p_2$ ($\varepsilon_2$) and $p_3$
($\varepsilon_3$) are the four-momenta (polarization vector) of the
initial state $\chi_{c1}^{\prime}$, final state $V_1$ and $V_2$,
respectively. $q_1$, $q_2$ and $q_3$ are the four-momenta of the up,
right and down charmed mesons in the triangle loop, respectively.

Similarly, the explicit transition amplitudes for
$\chi_{c2}^{\prime}(p_1)\to [D^{(*)}(q_1) {\bar D}^{(*)}(q_3)]
D^{(*)}(q_2) \to V(p_2)P(p_3)$ are given by

\begin{eqnarray}
\mathcal{M}_a&=& \int\frac{d^4q_2}{(2\pi)^4} [ig_{\chi_{c2}^{\prime}D^*D^*}\epsilon_{1\alpha\beta}] [2f_{D^*DV} \varepsilon_{\mu\nu\theta\phi} (q_1+q_2)^\theta ] \nonumber \\
&& \times p_2^\mu \epsilon_2^{*\nu} [g_{D^*DP}p_3^\kappa]\frac {i(-g^{\alpha\phi}+{q^\alpha_1 q^\phi_1}/{m_1^2})} {q_1^2-m_1^2} \nonumber \\
&&\times\frac {i}
 {q_2^2-m_2^2} \frac {i(-g^{\beta\kappa}+{q^\beta_3q^\kappa_3}/{m_3^2})} {q_3^2-m_3^2} {\cal F}(q^2), 
\end{eqnarray}

\begin{eqnarray}
\mathcal{M}_b&=& \int\frac{d^4q_2}{(2\pi)^4} [ig_{\chi_{c2}^{\prime}D^*D^*}\epsilon_{1\alpha\beta}] [g_{D^*D^*V} (q_1+q_2)_\phi g_{\mu\theta}\epsilon_2^{*\phi} \nonumber \\
&& -4f_{D^*D^*V}(p_{2\theta}g_{\mu\phi}-p_{2\mu}g_{\theta\phi})\epsilon_2^{*\phi}]\nonumber \\
&&\times[-\frac{1}{2}g_{D^*D^*P}\varepsilon_{\kappa\lambda\rho\sigma}p_3^\lambda q_2^\rho]\frac {i(-g^{\alpha\mu}+{q^\alpha_1q^\mu_1}/{m_1^2})} {q_1^2-m_1^2} \nonumber \\
&&\times \frac {i(-g^{\theta\kappa}+{q^\theta_2q^\kappa_2}/{m_2^2})}
 {q_2^2-m_2^2} \frac {i(-g^{\beta\sigma}+{q^\beta_3q^\sigma_3}/{m_3^2})} {q_3^2-m_3^2} {\cal F}(q^2) \, , \nonumber \\
\end{eqnarray}
with $\varepsilon_1$ the polarization tensor of initial state
$\chi_{c2}^{\prime}$.

In the triangle diagrams of Figs.~\ref{fig:feyn_chic1} and~\ref{fig:feyn_chic2}, the intermediate charmed mesons are usually off-shell. To compensate the off
shell effects and regularize the ultraviolet divergence~\cite{Li:1996yn,Locher:1993cc,Li:1996cj}, we adopt the following form factors,
\begin{eqnarray}\label{ELA-form-factor}
{\cal F}(q^2) \equiv \prod_i \left(\frac
{\Lambda_i^2-m_{i}^2} {\Lambda_i^2-q_i^2}\right),
\end{eqnarray}
where $i=1,2,3$ corresponds three intermediate mesons, respectively.
$\Lambda_i \equiv m_i+\alpha\Lambda_{\rm QCD}$ and the QCD energy
scale $\Lambda_{\rm QCD} = 220$ MeV. In the present work, the model
parameter $\alpha$ is constrained between $0.2$ and $0.4$ for
$\chi'_{c1} \to VV$ decays, and $0.4$ and $0.8$ for $\chi'_{c2} \to
VP$ decays. It is worth to mention that, with the values of $\alpha$
in the above range, the experimental data on the decays of
$\chi_{c1} \to VV$ and $\chi_{c2} \to VP$ can be well
reproduced~\cite{Liu:2009vv}.

\section{Numerical results} \label{sec:numerical-results}

In this section, we first determine the coupling constants in the
above section and then present our numerical results. Under the
heavy quark limit, the coupling constants of $P$-wave charmonium
coupled to the charmed mesons are as
follows~\cite{Casalbuoni:1996pg,Zhao:2013jza}:
\begin{eqnarray}
g_{\chi_{c1}^{\prime}{\cal D}^*{\cal D}} &=& 2\sqrt{2}g_1
\sqrt{m_{\chi_{c1}^{\prime}} m_{\cal D}
m_{{\cal D}^*}} \, , \\
g_{\chi_{c2}^{\prime}{\cal D}^*{\cal D}^*} &=& 4g_1 m_{{\cal
D}^*}\sqrt{m_{\chi_{c2}^{\prime}}} \, ,
\end{eqnarray}
with $g_1 =1.28$ GeV$^{-1/2}$, which is obtained in the linear
potential model~\cite{Deng:2016stx,Gui:2018rvv}. Besides, the
charmed meson couplings to the light vector mesons are obtained
with~\cite{Casalbuoni:1996pg,Cheng:2004ru},
\begin{eqnarray}
g_{{\cal D}{\cal D}V} &=& g_{{\cal D}^*{\cal D}^*V} = \frac{\beta g_V}{\sqrt{2}} , \\
f_{{\cal D}^*{\cal D}V} &=& \frac{ f_{{\cal D}^*{\cal D}^*V}}{m_{{\cal D}^*}}=\frac{\lambda g_V}{\sqrt{2}} \, , \\
g_{\mathcal{D}^{*} \mathcal{D} \mathcal{P}} &=& \frac{2 g}{f_{\pi}}
\sqrt{m_{\mathcal{D}} m_{\mathcal{D}^{*}}}, \\
g_{\mathcal{D}^{*} \mathcal{D}^{*} \mathcal{P}} &=& \frac{g_{{\cal
D}^{*} {\mathcal D} {\mathcal {P}}}}{\sqrt{m_{\mathcal{D}}
m_{\mathcal{D}^{*}}}} ,
\end{eqnarray}
with $g_V = {m_\rho / f_\pi}$ and $f_\pi = 132$
MeV~\cite{Wu:2016ypc}, $\beta=0.9$, $\lambda = 0.56 \, {\rm
GeV}^{-1} $ and $g=0.59$~\cite{Isola:2003fh}.

\subsection{$\chi'_{c1} \to VV$}

The $X(3872)$ resonance has the same quantum numbers as
$\chi_{c1}^{\prime}$ but with a much lighter mass than the one
predicted by potential quark model. Hence, we study the partial
decay widths of the $\chi'_{c1} \to VV$ as a function of the mass of
$\chi_{c1}^\prime$ from $3.8$ to $4.0$ GeV, which covers the
predicted values of the quark
models~\cite{Barnes:2005pb,Li:2009zu,Deng:2016stx}.

\begin{figure}[htbp]
\centering
\includegraphics[scale=0.5]{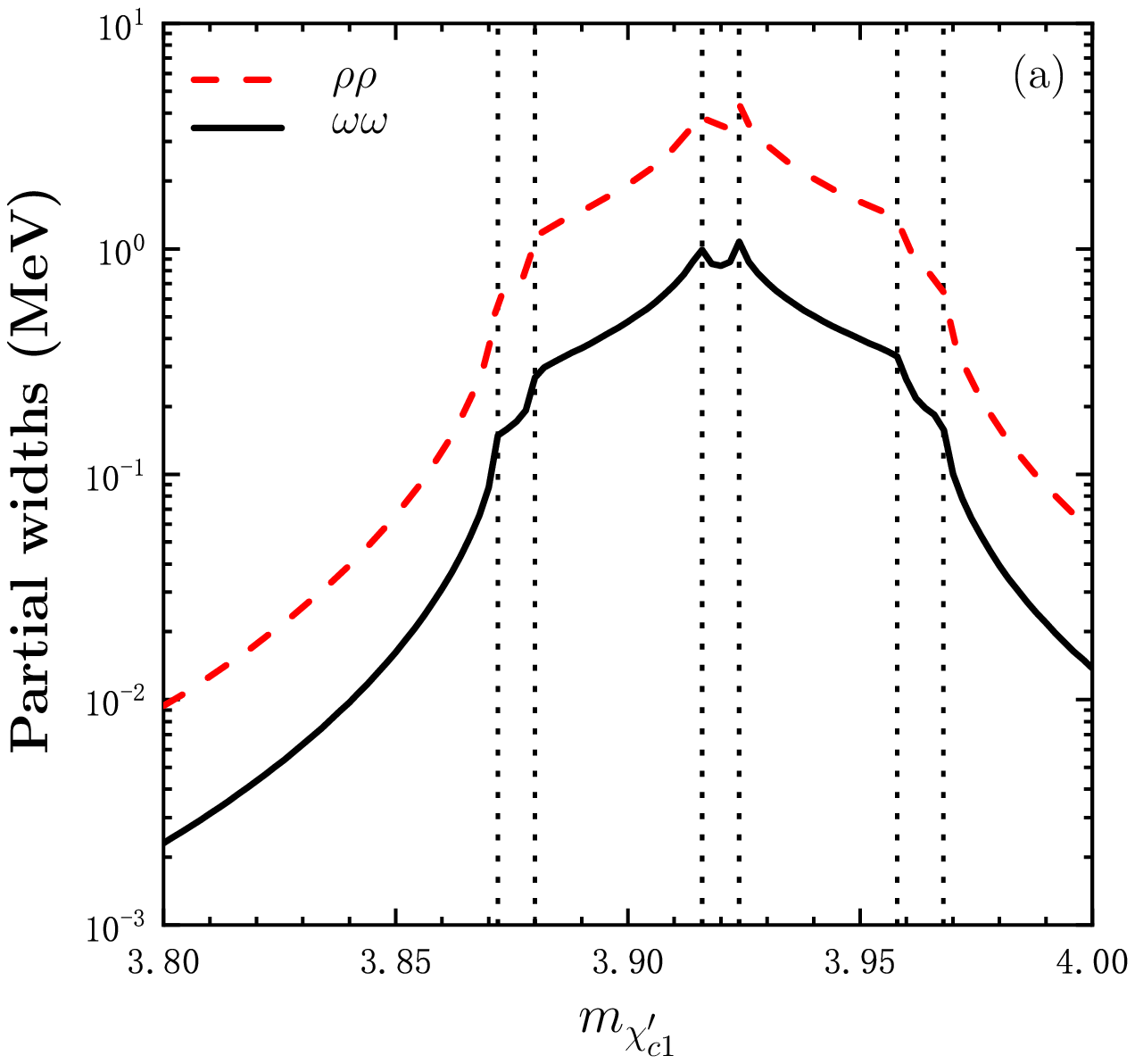} \\
\includegraphics[scale=0.5]{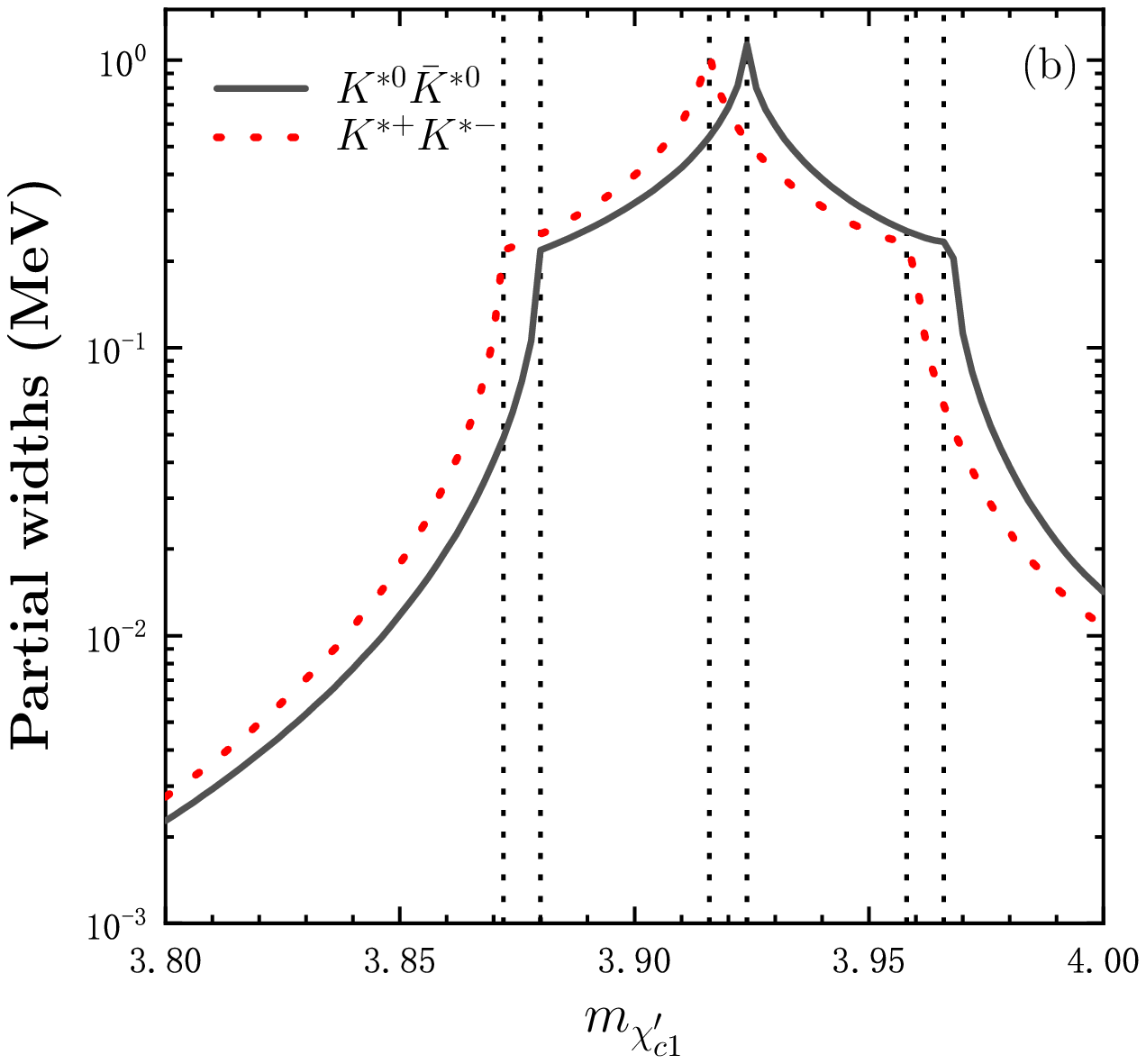} \\
\includegraphics[scale=0.5]{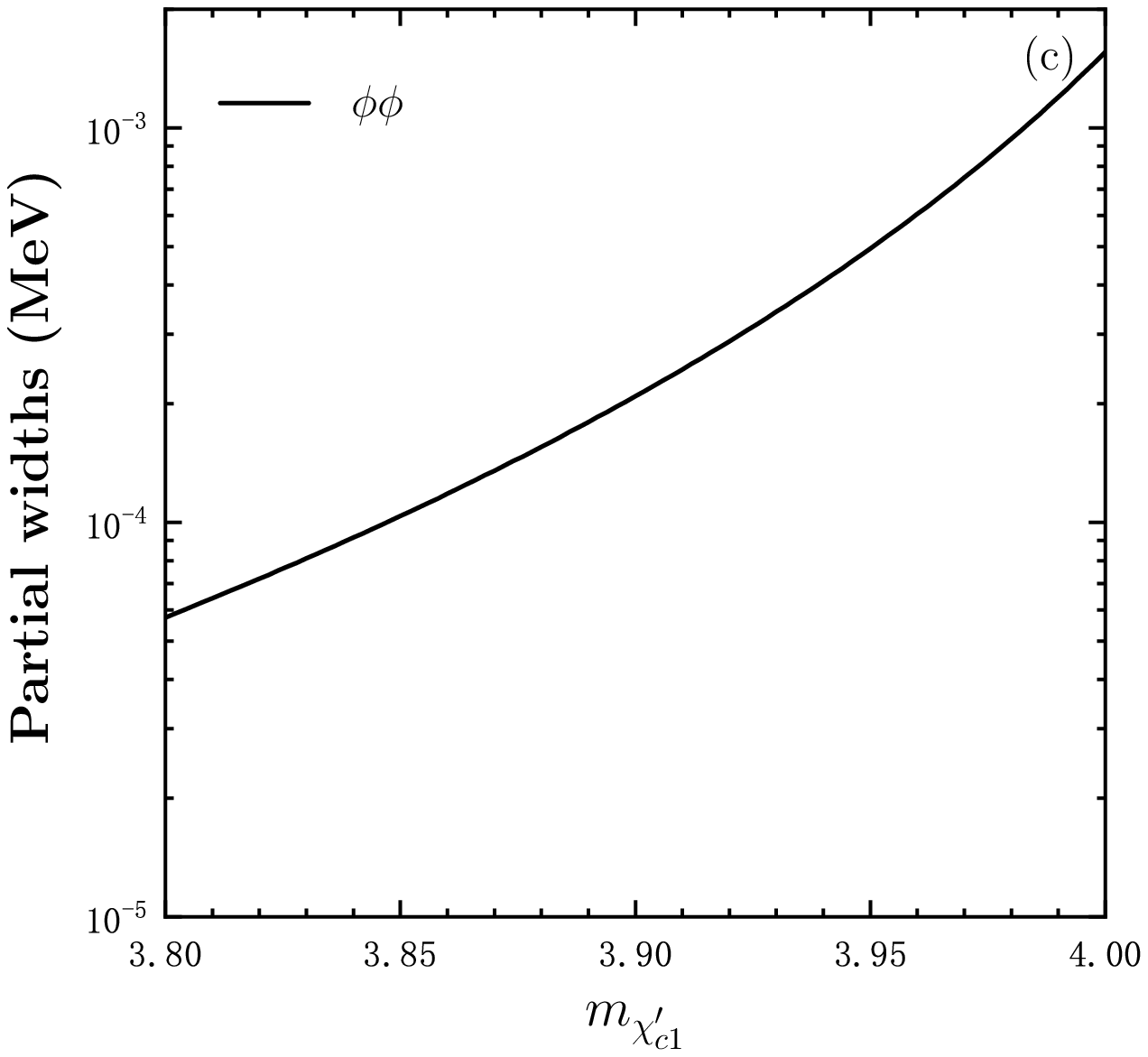}\\
\caption{$m_{\chi_{c1}^\prime}$-dependence of the decay width of
$\chi_{c1}^\prime \rightarrow VV$ with
$\alpha=0.2$.}\label{Fig:Tri1}
\end{figure}

In Fig.~\ref{Fig:Tri1}(a), we plot the partial widths of
$\chi_{c1}^{\prime} \to \omega\omega$ (solid line) and $\rho\rho$
(dashed line) in terms of the mass of $\chi_{c1}^\prime$ with
$\alpha=0.2$, respectively. From Fig.~\ref{Fig:Tri1}(a), one can see
that the calculated partial widths are sensitive to the mass of
$\chi_{c1}^\prime$, which can vary from order of keV to order of
MeV. Since these two decay modes have the same intermediate charmed
meson loops as shown in Fig.~\ref{fig:feyn_chic1}, they have the
similar behavior as a function of the mass of $\chi'_{c1}$. In
Fig.~\ref{Fig:Tri1}(b), we show the partial widths of $\chi'_{c1}
\to K^{*0} {\bar K}^{*0}$ (solid line) and $K^{*+} K^{*-}$ (dashed
line). For $\chi'_{c1} \to K^{*0} {\bar K}^{*0}$, the transition is
via $[D^+D^{*-}]D_s^{(*)}$ intermediate mesons in
Fig.~\ref{fig:feyn_chic1}, which leads to an enhancement at the
$D^+D^{*-}$ threshold. Similarly, the $\chi'_{c1} \to K^{*+} K^{*-}$
transition is via $[D^0 {\bar D}^{*0}]D_s^{(*)}$ intermediate
mesons, which leads to an enhancement at the $D^0 {\bar D}^{*0}$
threshold. In Fig.~\ref{Fig:Tri1}(c), we show our numerical results
for $\chi_{c1}^\prime \to \phi\phi$ decay, where there is no cusp
structure. This is because the mass range of $\chi_{c1}^\prime$ lies
below the intermediate $D_s^*{\bar D}_s^*$ threshold. The calculated
partial widths of $\chi_{c1}^\prime \to \phi\phi$ is about $3 \sim
4$ orders smaller than that of other decay modes in
Figs.~\ref{Fig:Tri1}(a) and (b). It indicates the threshold effects
via strange charmed meson pair is less important in
$\chi_{c1}^\prime \to \phi\phi$.

\begin{figure}[htbp]
\centering
\includegraphics[scale=0.5]{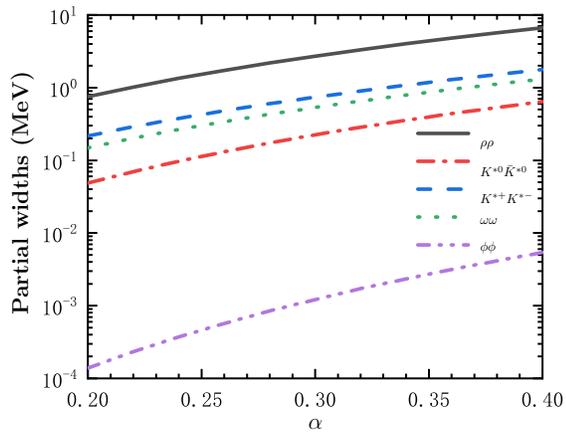}
\caption{$\alpha$-dependence of the decay width of
$\chi_{c1}^{\prime}\to VV$ with the $m_{\chi_{c1}^{\prime}}=3.872$
GeV.}\label{Fig:Tri1-mass}
\end{figure}

Next, we consider the form factors effects on the partial decay
widths with different cutoff parameters $\alpha$. In
Fig.~\ref{Fig:Tri1-mass}, the partial decay widths of
$\chi_{c1}^\prime \to VV$ are plotted in terms of $\alpha=0.2\sim
0.4$, and we take the mass of $\chi_{c1}^\prime = 3.872$ GeV. One
can see that the partial width of $\chi_{c1}^\prime \to \phi\phi$ is
about $2\sim 4$ orders smaller than the ones of other decay modes at
the same value of $\alpha$, which indicates that contribution of the
intermediate charmed meson loop to the process of $\chi_{c1}^\prime
\to \phi\phi$ is much smaller than the ones to other channels. This
is easily to be understand, since the mass of $\chi_{c1}^\prime$ is
much far away from the mass threshold of $D_s D_s^*$ than $DD^*$.

Furthermore, the total width of the $X(3872)$ is smaller than $1.2$
MeV as quoted in the PDG~\cite{Patrignani:2016xqp}. Some theoretical
works~\cite{Fleming:2007rp,Guo:2014hqa,Dai:2019hrf} suggest that the
width of $X(3872)$ should be less than $100$ keV based on the
molecule ansatz of $X(3872)$. The numerical results here as shown in
Fig.~\ref{Fig:Tri1-mass} is larger than the above upper limits of
$X(3872)$, which illustrate from the other side that the $X(3872)$
is at least not pure $c\bar c$ charmonium state, or there is only
small $c\bar c$ component in its wave function. We expect that the
more and precise experimental measurements on the relevant channels
can help us improving our understanding about the nature of the
$X(3872)$ state.

\subsection{$\chi'_{c2} \to VP$}

The $X(3930)$ was observed in the $\gamma \gamma \to D\bar D$
process by Belle~\cite{Uehara:2005qd} and Babar~\cite{Aubert:2010ab}
collaborations, and has been a good candidate for the
$\chi_{c2}^{\prime}$ state. The accessible decay modes are only
$\chi_{c2}^{\prime} \to K^*\bar K +c.c$ and $\rho^+\pi^- +c.c$ due
to $C$-parity conservation. In Tab.~\ref{Tab:br}, we present the
calculated partial widths of $\chi_{c2}^{\prime}$ for each channel
with cutoff parameters $\alpha=0.4$, $0.6$, and $0.8$, respectively.
As can be seen in Tab.~\ref{Tab:br}, the partial widths of
$\chi_{c2}^{\prime}\rightarrow K^\ast \bar{K}+c.c$ turn out to be
sizeable, while the partial widths of $\chi_{c2}^{\prime} \to
\rho^+\pi^- +c.c$ is found to be much smaller than the $K^\ast
\bar{K}+c.c$ channel. This is because of the $U$-spin symmetry
breaking caused by $u/d$ and $s$ quark mass difference in $K^\ast
\bar{K}+c.c$ channel is much larger than the isospin symmetry
breaking caused by $u$ and $d$ quark mass difference in $\rho\pi$
channel. If we take the $\chi_{c2}^{\prime}$ total width
$\Gamma_{\rm total}= 24$ MeV from PDG~\cite{Patrignani:2016xqp},
with the $\alpha=0.4 \sim 0.8$, the lower and upper bounds of ${\rm
BR}(\chi_{c2}^{\prime} \to K^* K +c.c)$ are about $1.25\times
10^{-3}$ and $1.9\%$, respectively. The lower and upper bounds of
${\rm BR}(\chi_{c2}^{\prime} \to \rho^+\pi^- +c.c)$ are about
$2.67\times 10^{-5}$ and $3.03\times 10^{-4}$, respectively.

\begin{table}
\begin{center}
\caption{The partial widths (in units of keV) of
$\chi_{c2}^\prime\rightarrow VP$ with different $\alpha$ values. We
take the mass of $\chi'_{c2}$ is 3927.2 MeV as in
PDG~\cite{Patrignani:2016xqp}. }\label{Tab:br}
\begin{tabular}{ccccccccc}
\hline \hline
Final states    & $\alpha=0.4$  & $\alpha=0.6$  & $\alpha=0.8$   \\
\hline
$\rho^\pm \pi^\mp$              & 0.32  & 1.40   & 3.63      \\
$K^{\ast0}\bar{K}^{0}+c.c.$     & 30.04 & 158.13 & 466.10    \\
$K^{\ast+}K^{-}+c.c.$           & 36.47 & 189.05 & 551.63    \\
\hline \hline
\end{tabular}
\end{center}
\end{table}

\begin{figure}[htb]
\centering
\includegraphics[width=0.45\textwidth]{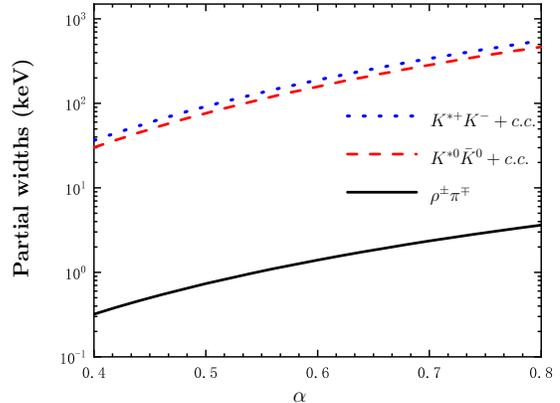}
\caption{$\alpha$-dependence of the decay width of
$\chi_{c2}^{\prime}\to VP$ with $m_{\chi_{c2}^{\prime}}=3.930$
GeV.}\label{Fig:Tri2}
\end{figure}

In Fig.~\ref{Fig:Tri2}, we plot the form factor parameter $\alpha$
dependence of the decay widths of $\chi_{c2}^{\prime}\rightarrow VP$
with $m_{\chi'_{c2}} = 3927.2$ MeV. In this work, we take a relative
smaller $\alpha$ range, between 0.4 and 0.8, which is acceptable
since the form factors for the off-shell effects arising from the
three intermediate mesons, instead of only the right exchanged meson
in the triangle loop.

In general we cannot provide the value of the cutoff parameter
$\alpha$, which should be determined by the experimental
measurements. But, it would be interesting to further clarify the
uncertainties arising from the introduction of form factors by
studying the cutoff parameter $\alpha$ dependence of the ratios
between different partial decay. For doing this, we next study the
ratios of different partial decay widths, where the effects of the
couplings of the vertexes are canceled, and these ratios only
reflects the open threshold effects through the intermediate charmed
meson loops. For the decays of $\chi_{c1}^\prime \to VV$, we define
the following ratios:
\begin{eqnarray}
R_1 &=& \frac {\Gamma(\chi_{c1}^\prime \to K^{*0} {\bar
K}^{*0})}{\Gamma(\chi_{c1}^\prime \to \rho\rho)} \, , \nonumber \\
R_2 &=& \frac {\Gamma(\chi_{c1}^\prime \to K^{*+} K^{*-})}{\Gamma(\chi_{c1}^\prime \to \rho\rho)} \, , \nonumber \\
R_3&=&\frac {\Gamma(\chi_{c1}^\prime \to
\omega\omega)}{\Gamma(\chi_{c1}^\prime \to \rho\rho)} \, , \nonumber
\\
R_4 &=& \frac {\Gamma(\chi_{c1}^\prime \to
\phi\phi)}{\Gamma(\chi_{c1}^\prime \to \rho\rho)} \, ,
\label{Eq:ratio-1}
\end{eqnarray}
and, for $\chi_{c2}^\prime \to VP$, we define,
\begin{eqnarray}
r_1 &=& \frac {\Gamma(\chi_{c2}^\prime \to K^{*+} K^-
+c.c.)}{\Gamma(\chi_{c2}^\prime \to \rho^+\pi^-)} \, , \nonumber \\
r_2 &=& \frac {\Gamma(\chi_{c2}^\prime \to K^{*0} {\bar
K}^{0}+c.c.)}{\Gamma(\chi_{c2}^\prime \to \rho^+\pi^-)} \, .
\label{Eq:ratio-2}
\end{eqnarray}

\begin{figure}[htbp]
\centering
\includegraphics[scale=0.45]{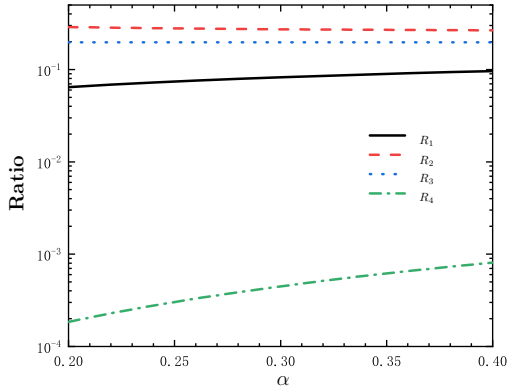}
\caption{The $\alpha$-dependence of the ratios defined in
Eq.~(\ref{Eq:ratio-1}) with $m_{\chi_{c1}^{\prime}}=3.872$
GeV.}\label{Fig:ratio-1}
\end{figure}

\begin{figure}[htbp]
\centering
\includegraphics[scale=0.45]{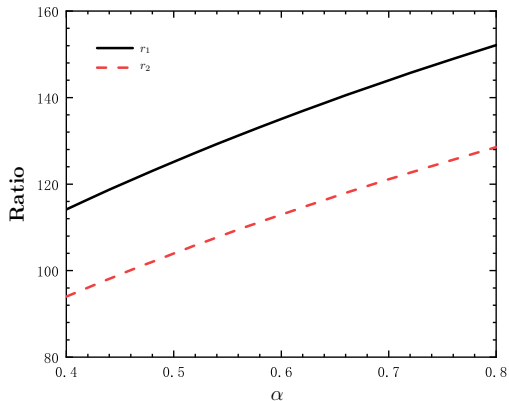}
\caption{The $\alpha$-dependence of the ratios defined in
Eq.~(\ref{Eq:ratio-2}) with $m_{\chi_{c2}^{\prime}}=3.930$
GeV.}\label{Fig:ratio-2}
\end{figure}

We show the numerical results for these ratios of
Eqs.~(\ref{Eq:ratio-1}) and (\ref{Eq:ratio-2}) in
Figs.~\ref{Fig:ratio-1} and \ref{Fig:ratio-2} as a function of the
cutoff parameter $\alpha$, from where we see that the dependence of
these ratios on the cutoff parameter $\alpha$ is rather weak, which
shows the validity of our model. These numerical results can be
tested by the experimental measurements in future.

\section{summary} \label{sec:summary}

In this work, we have investigated the charmless decays
$\chi_{c1}^\prime \to VV$ and $\chi_{c2}^\prime\to VP$ via IML,
which are supposed to be suppressed by HSR in pQCD. Applying an
effective Lagrangian approach, the charmed meson loop contributions
are calculated for evading HSR. We determined the cutoff parameter
$\alpha$ values for $\chi_{c1} \to VV$ and $\chi_{c2} \to VP$ by
reproducing the experimental data, which guide us to determine the
cutoff $\alpha$ range in $\chi_{c1}^\prime \to VV$ and
$\chi_{c2}^\prime \to VP$. Our results have shown that the
intermediate charmed meson loops may play an important role in these
processes, especially when the initial states are close to the two
particle thresholds.

For $\chi_{c1}^\prime \to VV$, the partial widths of $\rho\rho$,
$\omega\omega$, and $K^*{\bar K}^*$ channels can reach to order of
MeV, while partial widths of $\phi\phi$ channel can only reach to
order of keV. For $\chi_{c2}^\prime \to VP$, the partial widths of
$\chi_{c2}^{\prime}\rightarrow K^\ast \bar{K}+c.c$ turns out to be
sizeable, while the partial widths of $\chi_{c2}^{\prime} \to
\rho^+\pi^- +c.c$ is found to be much smaller than the $K^\ast
\bar{K}+c.c$ channel. This is because of the $U$-spin symmetry
breaking caused by $u/d-s$ quark mass difference in $K^\ast
\bar{K}+c.c$ channel is much larger than the isospin symmetry
breaking caused by $u-d$ quark mass difference in $\rho\pi$ channel.
Our calculations may be examined by the future BESIII experiments.

\section*{Acknowledgements}

We thank Xiao-Hai Liu for useful discussions. This work is supported
by the National Natural Science Foundation of China, under Grants
Nos. 11675091, 11835015, 11735003, 1191101015 and 11675131, the
Youth Innovation Promotion Association CAS (2016367) and the Higher Educational Youth Innovation Science and Technology Program Shandong Province £¨Grant No. 2020KJJ004£©


\begin{thebibliography}{0}
\bibitem{Chernyak:1981zz}
  V.~L.~Chernyak and A.~R.~Zhitnitsky,
  Nucl.\ Phys.\ B {\bf 201}, 492 (1982)  [Erratum-ibid.\ B {\bf 214}, 547 (1983)].

\bibitem{Chernyak:1983ej}
  V.~L.~Chernyak and A.~R.~Zhitnitsky,
  Phys.\ Rept.\  {\bf 112}, 173 (1984).

\bibitem{Feldmann:2000hs}
  T.~Feldmann and P.~Kroll,
  Phys.\ Rev.\ D {\bf 62}, 074006 (2000).

\bibitem{Lipkin:1988tg}
  H.~J.~Lipkin and S.~F.~Tuan,
  Phys.\ Lett.\ B {\bf 206}, 349 (1988).

\bibitem{Moxhay:1988ri}
  P.~Moxhay,
  Phys.\ Rev.\ D {\bf 39}, 3497 (1989).

\bibitem{Lipkin:1986bi}
  H.~J.~Lipkin,
  Nucl.\ Phys.\ B {\bf 291}, 720 (1987).

\bibitem{Lipkin:1986av}
  H.~J.~Lipkin,
  Phys.\ Lett.\ B {\bf 179}, 278 (1986).

\bibitem{Wang:2013hga}
Q.~Wang, C.~Hanhart and Q.~Zhao,
Phys.\ Lett.\ B \textbf{725}, 106 (2013).

\bibitem{Cleven:2013sq}
M.~Cleven, Q.~Wang, F.~-K.~Guo, C.~Hanhart,
U.~-G.~Mei{\ss }ner and Q.~Zhao,
Phys.\ Rev.\ D \textbf{87}, 074006 (2013).

\bibitem{Liu:2013vfa}
X.~-H.~Liu and G.~Li,
Phys.\ Rev.\ D \textbf{88}, 014013 (2013).

\bibitem{Guo:2013zbw}
F.~-K.~Guo, C.~Hanhart, U.~-G.~Mei{\ss }ner, Q.~Wang
and Q.~Zhao, 
Phys.\ Lett.\ B \textbf{725}, 127 (2013).

\bibitem{Voloshin:2013ez}
M.~B.~Voloshin,
Phys.\ Rev.\ D \textbf{87}, 074011 (2013). 

\bibitem{Voloshin:2011qa}
M.~B.~Voloshin,
Phys.\ Rev.\ D \textbf{84}, 031502 (2011). 

\bibitem{Li:2013zcr}
G.~Li, X.~h.~Liu, Q.~Wang and Q.~Zhao,
Phys.\ Rev.\ D \textbf{88}, 014010 (2013).

\bibitem{Li:2011ssa}
G.~Li and Q.~Zhao,
Phys.\ Rev.\ D \textbf{84}, 074005 (2011).

\bibitem{Chen:2011pv}
D.~-Y.~Chen and X.~Liu,
Phys.\ Rev.\ D \textbf{84}, 094003 (2011). 

\bibitem{Li:2013yla}
G.~Li and X.~-H.~Liu,
Phys.\ Rev.\ D \textbf{88}, 094008 (2013). 

\bibitem{Chen:2011pu}
D.~-Y.~Chen, X.~Liu and T.~Matsuki,
Phys.\ Rev.\ D \textbf{84}, 074032 (2011). 

\bibitem{Chen:2012yr}
  D.~Y.~Chen, X.~Liu and T.~Matsuki,
  Chin.\ Phys.\ C {\bf 38}, 053102 (2014).

\bibitem{Bondar:2011ev}
A.~E.~Bondar, A.~Garmash, A.~I.~Milstein, R.~Mizuk
and M.~B.~Voloshin, 
Phys.\ Rev.\ D \textbf{84}, 054010 (2011). 

\bibitem{Li:2015uwa}
G.~Li and Z.~Zhou,
Phys.\ Rev.\ D \textbf{91}, 034020 (2015). 

\bibitem{Li:2014gxa}
 G.~Li, C.~S.~An, P.~Y.~Li, D.~Liu, X.~Zhang and
Z.~Zhou, 
Chin.\ Phys.\ C \textbf{39}, 063102 (2015). 

\bibitem{Chen:2013bha}
D.~-Y.~Chen, X.~Liu and T.~Matsuki,
Phys.\ Rev.\ D \textbf{88}, 014034 (2013). 

\bibitem{Li:2012as}
G.~Li, F.~l.~Shao, C.~W.~Zhao and Q.~Zhao,
Phys.\ Rev.\ D \textbf{87}, 034020 (2013).

\bibitem{Li:2014uia}
G.~Li and W.~Wang,
Phys.\ Lett.\ B \textbf{733}, 100 (2014).

\bibitem{Guo:2010ak}
F.~K.~Guo, C.~Hanhart, G.~Li, U.~G.~Meissner and
Q.~Zhao, 
Phys.\ Rev.\ D \textbf{83}, 034013 (2011) 

\bibitem{Wu:2016ypc}
  Q.~Wu, G.~Li, F.~Shao and R.~Wang,
  Phys.\ Rev.\ D {\bf 94}, 014015 (2016).

\bibitem{Wu:2016dws}
  Q.~Wu, G.~Li, F.~Shao, Q.~Wang, R.~Wang, Y.~Zhang and Y.~Zheng,
  Adv.\ High Energy Phys.\  {\bf 2016}, 3729050 (2016).

\bibitem{Liu:2016xly}
  X.~H.~Liu and G.~Li,
  Eur.\ Phys.\ J.\ C {\bf 76}, 455 (2016).

\bibitem{Li:2014pfa}
  G.~Li, X.~H.~Liu and Z.~Zhou,
  Phys.\ Rev.\ D {\bf 90}, 054006 (2014).

\bibitem{Yuan-Jiang:2010cna}
  Y.~J.~Zhang, G.~Li and Q.~Zhao,
  Chin.\ Phys.\ C {\bf 34}, 1181 (2010).

\bibitem{Zhao:2013jza}
  C.~W.~Zhao, G.~Li, X.~H.~Liu and F.~L.~Shao,
  Eur.\ Phys.\ J.\ C {\bf 73}, 2482 (2013).

\bibitem{Li:2013xia}
  G.~Li,
  Eur.\ Phys.\ J.\ C {\bf 73}, 2621 (2013).

\bibitem{Li:2007xr}
  G.~Li and Q.~Zhao,
  Phys.\ Lett.\ B {\bf 670}, 55 (2008).

\bibitem{Qin:2019ybr}
  W.~H.~Qin, C.~S.~An, G.~Li, C.~Wang and Y.~Wang,
  Eur.\ Phys.\ J.\ C {\bf 79}, 757 (2019).

\bibitem{Liu:2019dqc}
  X.~H.~Liu, G.~Li, J.~J.~Xie and Q.~Zhao,
  Phys.\ Rev.\ D {\bf 100}, 054006 (2019).

\bibitem{Wu:2019vbk}
  Q.~Wu, D.~Y.~Chen, X.~J.~Fan and G.~Li,
  Eur.\ Phys.\ J.\ C {\bf 79}, 265 (2019).

\bibitem{Zhang:2018eeo}
  Y.~Zhang and G.~Li,
  Phys.\ Rev.\ D {\bf 97}, 014018 (2018).

\bibitem{Patrignani:2016xqp}
  C.~Patrignani {\it et al.} [Particle Data Group],
  Chin.\ Phys.\ C {\bf 40}, 100001 (2016).

\bibitem{Liu:2010um}
  X.~H.~Liu and Q.~Zhao,
  J.\ Phys.\ G {\bf 38}, 035007 (2011).

\bibitem{Wang:2012wj}
  Q.~Wang, X.~H.~Liu and Q.~Zhao,
  Phys.\ Lett.\ B {\bf 711}, 364 (2012).

\bibitem{Liu:2009vv}
  X.~H.~Liu and Q.~Zhao,
  Phys.\ Rev.\ D {\bf 81}, 014017 (2010).

\bibitem{Li:2013jma}
G.~Li, X.~H.~Liu and Q.~Zhao,
Eur.\ Phys.\ J.\ C \textbf{73}, 2576 (2013).

\bibitem{Wang:2012mf}
  Q.~Wang, G.~Li and Q.~Zhao,
  Phys.\ Rev.\ D {\bf 85}, 074015 (2012).

\bibitem{Casalbuoni:1996pg}
  R.~Casalbuoni, A.~Deandrea, N.~Di Bartolomeo, R.~Gatto, F.~Feruglio and G.~Nardulli,
  Phys.\ Rept.\  {\bf 281}, 145 (1997).

\bibitem{Colangelo:2003sa}
  P.~Colangelo, F.~De Fazio and T.~N.~Pham,
  Phys.\ Rev.\ D {\bf 69}, 054023 (2004).

\bibitem{Cheng:2004ru}
  H.~Y.~Cheng, C.~K.~Chua and A.~Soni,
  Phys.\ Rev.\ D {\bf 71}, 014030 (2005).

\bibitem{Li:1996yn}
  X.~-Q.~Li, D.~V.~Bugg and B.~-S.~Zou,
  Phys.\ Rev.\ D {\bf 55}, 1421 (1997).

\bibitem{Locher:1993cc}
  M.~P.~Locher, Y.~Lu and B.~S.~Zou,
  Z.\ Phys.\ A {\bf 347}, 281 (1994).

\bibitem{Li:1996cj}
  X.~-Q.~Li and B.~-S.~Zou,
  Phys.\ Lett.\ B {\bf 399}, 297 (1997).

\bibitem{Deng:2016stx}
  W.~J.~Deng, H.~Liu, L.~C.~Gui and X.~H.~Zhong,
  Phys.\ Rev.\ D {\bf 95}, 034026 (2017).

\bibitem{Gui:2018rvv}
  L.~C.~Gui, L.~S.~Lu, Q.~F.~Lv, X.~H.~Zhong and Q.~Zhao,
  Phys.\ Rev.\ D {\bf 98}, 016010 (2018).

\bibitem{Isola:2003fh}
  C.~Isola, M.~Ladisa, G.~Nardulli and P.~Santorelli,
  Phys.\ Rev.\  D {\bf 68}, 114001 (2003).

\bibitem{Li:2009zu}
  B.~Q.~Li and K.~T.~Chao,
  Phys.\ Rev.\ D {\bf 79}, 094004 (2009).

\bibitem{Barnes:2005pb}
  T.~Barnes, S.~Godfrey and E.~S.~Swanson,
  Phys.\ Rev.\ D {\bf 72}, 054026 (2005).

\bibitem{Fleming:2007rp}
S.~Fleming, M.~Kusunoki, T.~Mehen and U.~van Kolck,
Phys.\ Rev.\ D \textbf{76}, 034006 (2007)

\bibitem{Guo:2014hqa}
F.~Guo, C.~Hidalgo-Duque, J.~Nieves, A.~Ozpineci and M.~P.~Valderrama,
Eur.\ Phys.\ J.\ C \textbf{74}, 2885 (2014).

\bibitem{Dai:2019hrf}
L.~Dai, F.~Guo and T.~Mehen,
Phys.\ Rev.\ D \textbf{101}, no.5, 054024 (2020)

\bibitem{Uehara:2005qd}
  S.~Uehara {\it et al.} [Belle Collaboration],
  Phys.\ Rev.\ Lett.\  {\bf 96}, 082003 (2006).

\bibitem{Aubert:2010ab}
  B.~Aubert {\it et al.} [BaBar Collaboration],
  Phys.\ Rev.\ D {\bf 81}, 092003 (2010).
\end{thebibliography}
\end{document}